\documentclass{article}
\usepackage{spconf,amsmath,graphicx}
\usepackage{multirow}
\usepackage{booktabs}

\newcommand{\MB}[1]{\mbox{\boldmath{$#1$}}}

\begin{document}

\name{
\parbox{.95\linewidth}{\centering
Khe Chai Sim, Fran{\c{c}}oise Beaufays, Arnaud Benard, Dhruv Guliani, Andreas Kabel, Nikhil Khare, Tamar Lucassen, Petr Zadrazil, Harry Zhang, Leif Johnson, Giovanni Motta, Lillian Zhou}}
\address{Google, USA}

\title{Personalization of End-to-end Speech Recognition On Mobile Devices For Named Entities}

\maketitle

\begin{abstract}
We study the effectiveness of several techniques to personalize end-to-end speech models and improve the recognition of proper names relevant to the user. These techniques differ in the amounts of user effort required to provide supervision, and are evaluated on how they impact speech recognition performance. 
We propose using {\em keyword-dependent} precision and recall metrics to measure vocabulary acquisition performance. 
We evaluate the algorithms on a dataset that we designed to contain names of persons that are difficult to recognize. Therefore, the baseline recall rate for proper names in this dataset is very low: 2.4\%. A data synthesis approach we developed brings it to 48.6\%, with no need for speech input from the user. With speech input, if the user corrects only the names, the name recall rate improves to 64.4\%. If the user corrects all the recognition errors, we achieve the best recall of 73.5\%. To eliminate the need to upload user data and store personalized models on a server, we focus on performing the entire personalization workflow on a mobile device. 
\end{abstract}
\section{Introduction}
\label{sec:intro}

End-to-end speech recognition systems based on the Recurrent Neural Network Transducer architecture (RNN-T)~\cite{graves2012sequence} have been shown to achieve state-of-the-art performance for large-vocabulary continuous speech recognition~\cite{rao2017rnnt}. A RNN-T model has also been successfully deployed to run efficiently on mobile devices~\cite{he2019streaming}.
The ability to perform efficient on-device speech recognition opens up new avenues to perform on-device personalization of these models without the need for complex server-side infrastructure. More importantly, personalization means that user data and models are stored on users' devices and not sent to a centralized server, thus increasing data privacy and security.

To personalize speech models on device, we need a learning algorithm that is memory efficient and robust to overfitting. 
We also need users to provide labeled data, which can be time consuming.
To minimize the user effort needed, we investigate ways of improving on-device learning and compare personalization techniques that require different levels of user engagement.

The remainder of this paper is organized as follows.
Section~\ref{sec:personalization} describes the on-device speech personalization framework.
Section~\ref{sec:vocab_acquisition} presents several personalization methods that aim at learning new named entities.
In Section~\ref{sec:dataset}, we describe the Wiki-Names dataset, which we have collected specifically to evaluate the effectiveness of personalization techniques for learning new named entities.
Section~\ref{sec:eval} describes the evaluation metrics we use throughout our experiments.
Section~\ref{sec:results} presents the experimental results.

\section{On-device Speech Personalization}
\label{sec:personalization}

\begin{figure}[t]
\centering
\includegraphics[width=0.45\textwidth]{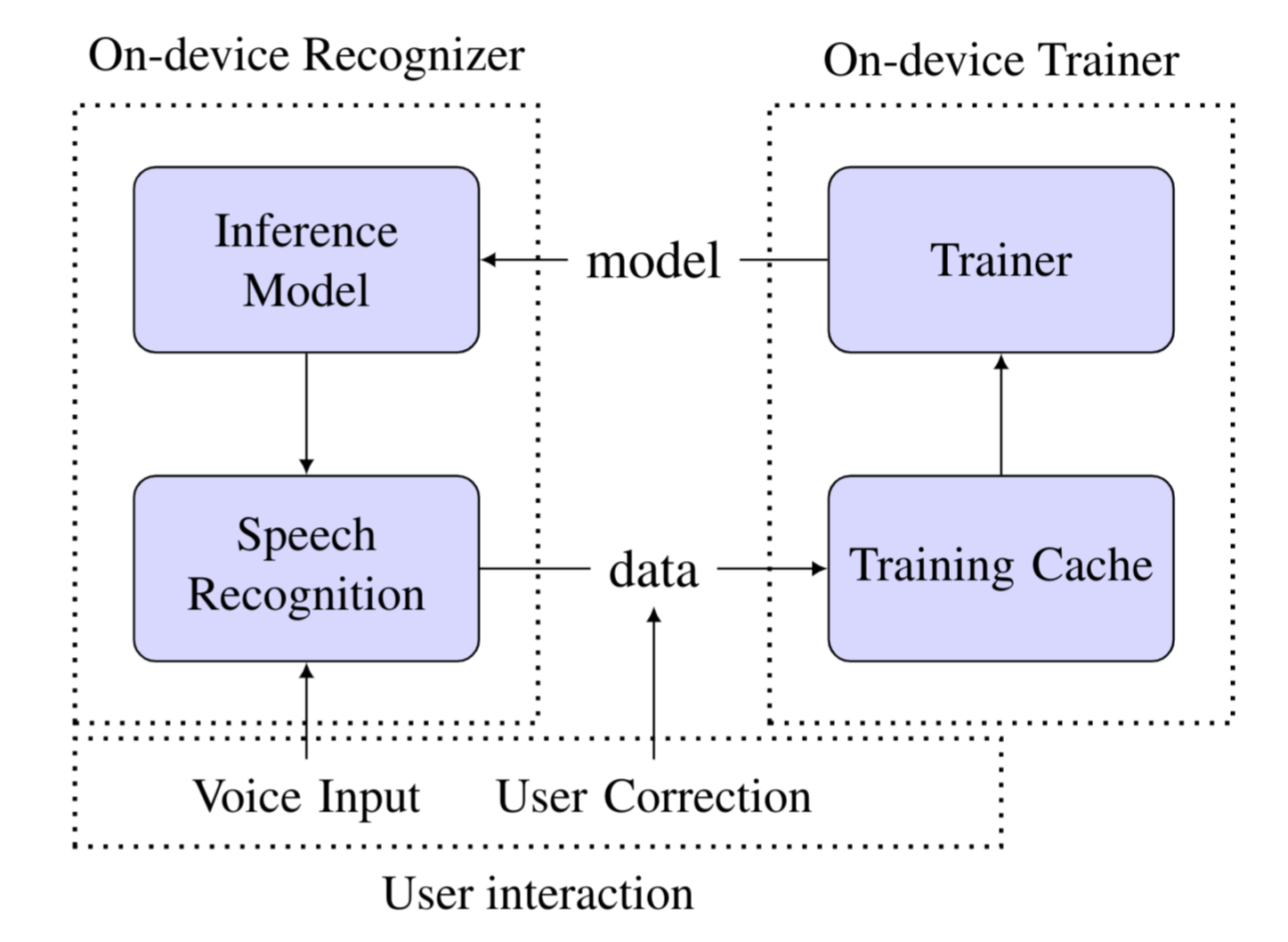}
\caption{On-device Speech Personalization Workflow.}
\label{fig:workflow}
\end{figure}
Fig.~\ref{fig:workflow} shows how the personalization workflow runs on a mobile device. It consists of two main phases: first, the user interacts with the device by voice. The user's speech is captured and transcribed into text using an on-device speech recognition system. The user may then optionally edit the recognition outputs to correct errors.
The input speech and corrected texts are stored in a {\em training cache} on device.
During the second phase, data from the training cache is used to perform on-device personalization of the speech model. This phase is usually performed while the device is idle so as to maximize the available memory and compute resources for personalization and preserve user experience. 
These two phases may take place in an interleaving fashion to achieve an online personalization process, where the system continuously adapts to the user's voice and usage patterns.

\begin{figure}[t]
\centering
\includegraphics[width=0.45\textwidth]{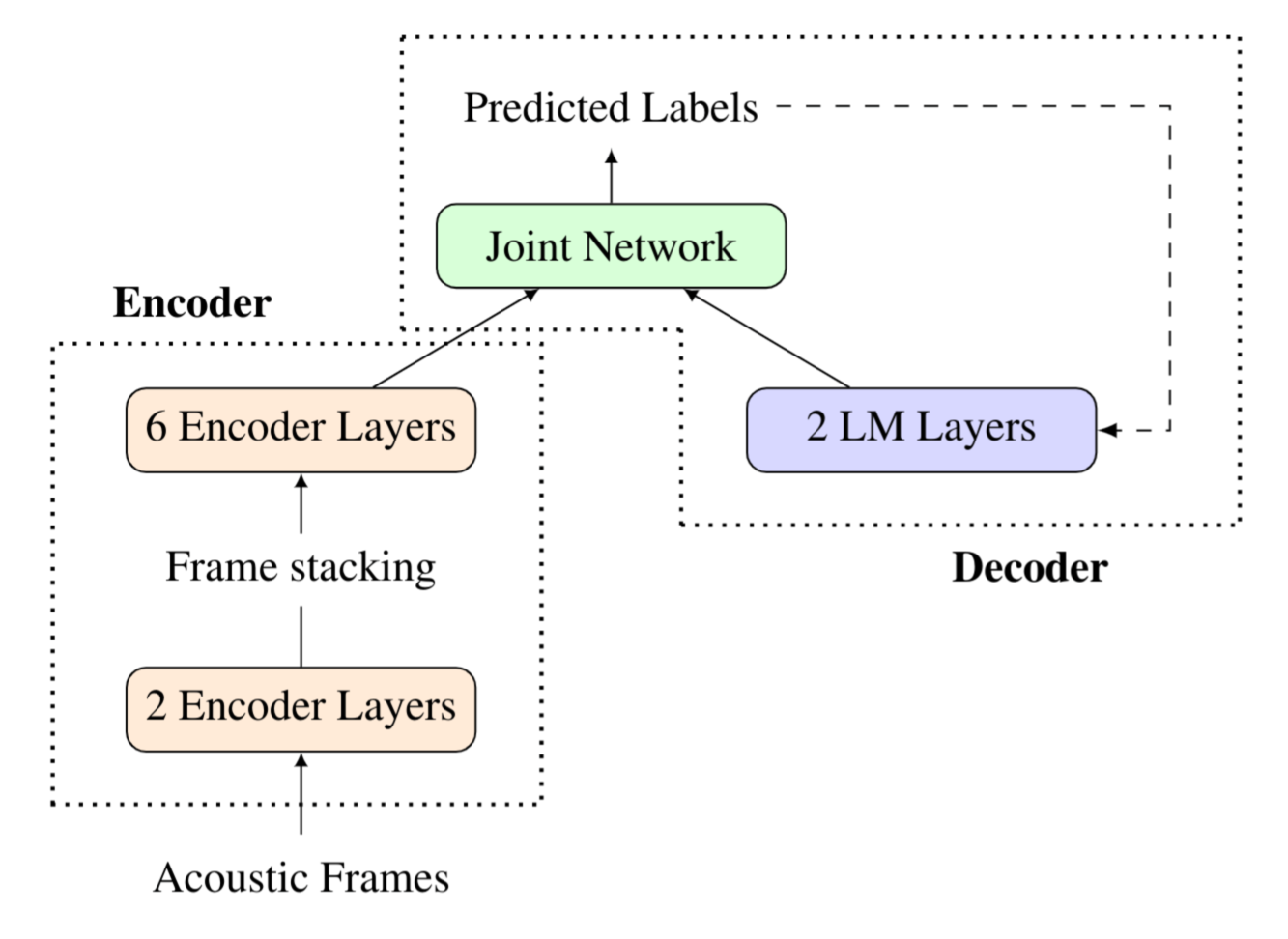}
\caption{Architecture of the on-device RNN-T model (8 encoder layers, 2 LM layers and a joint network with 1 hidden layer).}
\label{fig:rnnt_arch}
\end{figure}
Fig.~\ref{fig:rnnt_arch} depicts the RNN-T model architecture that will be used for the experiments in this paper. 
The model consists of an encoder with 8 Long Short Term Memory (LSTM)~\cite{sak2014long} layers for the acoustic features, an encoder with 2 LSTM layers for the label sequences (denoted as the language model (LM) component) and a joint network with a single hidden layer. The LM and joint network components are referred to as the decoder.
Each LSTM layer has 2048 hidden units along with a projection to 640 units.
3 consecutive frames of 80-dimensional log Mel features (extracted every 10 milliseconds) are stacked together to form the 240-dimensional inputs to the network~\cite{pundak2016lfr}.
There is also frame stacking after the second encoder layer with a stride size of 2 to increase the frame shift to 60 milliseconds.
The network outputs correspond to 75 graphemes and a blank symbol~\cite{graves2012sequence}.

\section{Vocabulary Acquisition}
\label{sec:vocab_acquisition}

Vocabulary acquisition is an important aspect of personalizing speech models, providing the system with the flexibility to learn new words that do not appear in the training data. The model becomes attuned to the small set of words that matter to the individual user, such as the names of family and friends.  

For an end-to-end sequence model, where the output units are sub-words, the vocabulary of the system is not explicitly defined. If the sub-word units are sufficiently defined to construct all possible words, there are technically no OOV words. However, words that do not appear in the training data will typically be assigned a low probability by the model, leading to recognition errors for rare or unseen words. 
The output probabilities of the new words can be increased by modifying the decoder to incorporate a biased language model~\cite{toshniwal2018comparison}. This includes techniques such as biasing~\cite{hall2015composition,aleksic2015improved,mcgraw2016personalized}, shallow fusion~\cite{gulcehre2015using}, deep fusion~\cite{gulcehre2015using} and cold fusion~\cite{sriram2017cold}.
Alternatively, the end-to-end model can also be fine-tuned to learn the new words.
In the following, we will describe the biasing and model fine-tuning approaches for handling new words. 

\subsection{Biasing}
\label{sec:biasing}

Biasing is a technique whereby new vocabulary is injected into the speech recognition system by boosting the language model probability of the new words during decoding. This approach does not require retraining of the existing model and can be viewed as a form of language model interpolation. This technique requires only a list of biasing words or phrases. 
It has been shown in~\cite{mcgraw2016personalized} that personalization can be achieved efficiently through new vocabulary injection and biasing the language model on the fly~\cite{hall2015composition,aleksic2015improved}.

\subsection{Model fine-tuning}
\label{sec:fine-tune}

A more direct way to learn new words is to fine-tune the model with training data that contains examples of the new words. This requires both speech and the corresponding transcription.
High-quality speech transcription is important to train speech recognition models.
For server-side training, this is typically outsourced to professional transcribers.
However, on-device personalization has to rely solely on individual users to provide training supervision. This would require a substantial amount of user effort and may not be practical for on-device training.

There are several ways to reduce user effort.
For example, users may provide a list of named entities important to them, or simply grant access to an existing list, such as contact names, place names and media names used by other apps.
A text-to-speech engine can then be used to synthesize the corresponding speech. This requires the same amount of user effort as biasing ({\em cf.} Section~\ref{sec:biasing}).
Alternatively, users may provide example sentences that contain the words of interest.
This will provide more contextual information for the new words and may improve generalization to unseen contexts.
Finally, users may also speak to the system and correct any recognition errors, as illustrated in the personalization workflow in Fig.~\ref{fig:workflow}.
If the user corrects every recognition error, we can achieve {\em supervised} learning. 
Otherwise, if the user only corrects some of the errors, we can perform {\em semi-supervised} learning.

\subsubsection{Overfitting}
\label{sec:over}

Although the speech RNN-T model has about 117M parameters, our results show that it is possible to fine-tune the entire model and achieve good improvements.
However, fine-tuning the full model can easily lead to overfitting. 
Overfitting can be suppressed by using a lower learning rate and early stopping. 
This is consistent with the results reported in~\cite{sim2018domain}, where robust domain adaptation can be achieved by fine-tuning the entire model with millions of parameters on a small amount of adaptation data. 

There are existing speaker adaptation techniques for neural networks that offer a more compact model representation by having a large number of global parameters shared across different speakers and a smaller number of speaker-specific parameters~\cite{sim2017_springer,li2010comparison,saon2013speaker,senior2014improving,swietojanski2014learning,tan2015cluster,wu2015multi,samarakoon2016factorized}.
These methods are more suitable for a centralized adaptation solution where the adapted models for all speakers are stored on a server.

On the other hand, regularization~\cite{srivastava2014dropout,kirkpatrick2016} and data augmentation~\cite{ko2015audio,lippmann1987multi} techniques can also be used to prevent overfitting when training on a small amount of data. They achieve robust model adaptation without modifying the model structure. These approaches are suitable for on-device personalization since the personalized models are stored separately on individual user devices. 
In the following section, we will describe the Elastic Weight Consolidation (EWC) regularization technique to mitigate the overfitting problem.

\subsubsection{Elastic Weight Consolidation}
\label{sec:ewc}

To perform personalization, we first train the model on a multi-speaker corpus, then fine-tune it on a single user. During the second training step, the model ``forgets'' about the previous task; this problem is called ``Catastrophic Forgetting''~\cite{mccloskey1989catastrophic}. We can mitigate this problem using Elastic Weight Consolidation (EWC)~\cite{kirkpatrick2016}, which aims at maximizing $\log p(\MB{\theta}|{\cal D})$,
or finding the most probable parameter values, $\MB{\theta}$, given some dataset, ${\cal D}$. By assuming that ${\cal D}$ contains data from two tasks, ${\cal D_A}$ and ${\cal D_B}$, and the model has already been trained on task A, the conditional probability is given by:
\begin{equation*}
    \log p(\MB{\theta}|{\cal D}) = \log p({\cal D_B}|\MB{\theta}) + \log p(\MB{\theta}|{\cal D_A}) - \log p(\MB{\theta}|{\cal D_B})
\end{equation*}
Note that $L_B(\MB{\theta}) = -\log p({\cal D_B}|\MB{\theta})$ is the standard loss function for task B. $\log p(\MB{\theta}|{\cal D_A})$ can be approximated as a Gaussian distribution with the mean given by
the parameters fine-tuned on task A and the precision matrix given by the Fisher Information Matrix, $F_i$. 
Therefore, the EWC loss function has an additional regularization term:
\[L(\MB{\theta}) = L_B(\MB{\theta}) + \dfrac{\lambda}{2}\sum_{i} F_i \times (\theta_i - \theta_{A,i}^{*})^2\]
where 
$\MB{\theta} = \{\theta_i\}$ denotes the set of model parameters. 
$\theta_{A,i}^{*}$ is the value of the $i$th parameter fine-tuned for task A.
$\lambda$ is the penalty weight that can be adjusted to control the amount of regularization.
The Fisher Information Matrix, $F_i$, is used to penalize moving in the directions critical (with high Fisher Information) to the first task. We compute $F_i$ by summing the square gradients of the loss with a model trained on the first task, as described in~\cite{kirkpatrick2016}.

\section{Dataset}
\label{sec:dataset}

We collected a dataset called {\em Wiki-Names} to evaluate the performance of speech personalization. These were sentences extracted from English Wikipedia pages containing repeated occurrences of politician and artist names that are difficult to recognize. To identify the difficult words, we extracted a list of names and ran our baseline speech recognition system on the synthesized speech for these names. The ones that were incorrectly recognized are deemed to be difficult words.
Four categories of names were chosen: {\em American}, {\em Chinese}, {\em Indian} and {\em Italian}. We recruited 100 participants to read the extracted sentences:
each speaker participated in one category only and were comfortable reading the names in their respective category. Nevertheless, the prompts were still difficult to read due to the presence of many name entities.
As a consequence, the resulting speech was also often accented and disfluent (with hesitations and corrections).

\begin{table}[t]
    \centering
    \caption{Wiki-Names Data}
    \label{tab:wiki_names}
    \begin{tabular}{c|c|c|c}
    \toprule
        \multirow{2}*{\textbf{Category}} & 
        \multirow{1}*{\textbf{No. of}} & 
        \multicolumn{2}{c}{\textbf{Amount of data (mins)}} \\
        & \textbf{Speakers} & \textbf{~~~Train~~~} & \textbf{Test} \\
    \midrule
        American & 6 & 24.7 & 9.7 \\
        Chinese & 24 & 126.0 & 49.7 \\
        Indian & 40 & 161.3 & 66.1 \\
        Italian & 30 & 144.6 & 57.9 \\
    \bottomrule
    \end{tabular}
\end{table}
Table~\ref{tab:wiki_names} shows the distribution of the data over the four categories.
Each participant provided 50 utterances (4.6 minutes) of training data and 20 utterances (1.9 minutes) of test data.
The prompts for each user covered five names, each with 10 training utterances and 4 test utterances, with each name potentially appearing multiple times per utterance. We manually transcribed the data to use as ground truth for both training and evaluation. 

\section{Evaluation}
\label{sec:eval}

We evaluate performance based on word error rate (WER), a standard metric used to evaluate speech recognition systems. 
In order to measure the performance based on a selected subset of words, we treat our problem as a retrieval task (similar to keyword spotting~\cite{zhang2009unsupervised}). 
We propose using {\em keyword-dependent} precision and recall to measure vocabulary acquisition performance, which can be computed as follows:
\begin{eqnarray}
    \text{precision} = \frac{\text{retrieved} \cap \text{relevant}}{\text{retrieved}} &=& \frac{N_c}{N_h} \\
    \text{recall} = \frac{\text{retrieved} \cap \text{relevant}}{\text{relevant}} &=& \frac{N_c}{N_r}
\end{eqnarray}
where $N_r$ and $N_h$ are the number of keywords in the reference and hypothesis, respectively. $N_c$ is the number of keywords that are correctly recognized.
To compute $N_c$, we first perform edit distance alignment between the reference and hypothesis texts. Then, we count the number of correct matches for the keywords only.

Precision and recall can better explain the performance of a personalization method. 
A low recall indicates the model's lack of ability to recognize the new entities 
({\em miss}), while a low precision indicates over-generation of those name entities ({\em false alarm}). 
\begin{table}[t]
    \centering
    \caption{An example of aligned reference and hypothesis texts. The relevant words are in bold face ($N_r = 3$, $N_h = 2$ and $N_c = 1$. Precision = $1/2$ and recall = $1/3$).}
    \label{tab:ref_hyp}
    \begin{tabular}{c||c|c|c|c|c|c}
        \toprule
         REF & {\bf Zhuge} & {\bf Dan} & was & from & {\bf Yangdu} &  \\
         HYP & {\bf Zhuge} &   & was & from & young & {\bf Zhuge}  \\
        \bottomrule
    \end{tabular}
\end{table}
For example, Table~\ref{tab:ref_hyp} shows a pair of aligned reference and hypothesis texts.
There are 3 relevant words in the reference (`Zhuge', `Dan' and `Yangdu') and 2 relevant words in the hypothesis (`Zhuge' and `Zhuge'). Only one keyword in the hypothesis has been aligned correctly with the reference. The recall is penalized by the system's inability to recognize `Dan' and `Yangdu' correctly while the precision is penalized by falsely recognizing `Zhuge' at the end.

Generally, as we train our models to learn new words, we observed that the recall will improve gradually as the model begins to recognize those new words. However, the personalized model also starts to over-generate those words in the output, which reduces precision. Therefore, precision and recall offer a way of measuring the trade-off.

\section{Experimental Results}
\label{sec:results}

In this section, we present experimental results to evaluate various techniques for personalizing the RNN-T speech recognition model. The architecture of this model is described in Section~\ref{sec:fine-tune}.
This model was trained with 35 million anonymized hand-transcribed English utterances (∼27,500 hours), from Google's voice search traffic~\cite{he2019streaming}.
All models are trained and evaluated using Tensorflow~\cite{abadi2016tensorflow}. The RNN-T loss and gradients are computed using the efficient implementation as described in~\cite{bagby2018efficient}. We fine-tune the models using the momentum optimizer~\cite{sutskever2013importance}
for 15 epochs with a learning rate of $10^{-4}$ and batch size of 5.

\subsection{Fine-tune Selected Layers}

\begin{table}[t]
    \centering
    \caption{Comparison of word error rate performance on Wiki-Names dataset (15 epochs).}
    \label{tab:wer_selected_layers}
    \begin{tabular}{l|c|rr}
    \toprule
        \multirow{2}*{\textbf{Model}} & 
        \multirow{1}*{\textbf{\# of fine-tuning}} & 
        \multicolumn{2}{c}{\textbf{WER}} \\
        & \textbf{Parameters} & \textbf{TTS} & \textbf{Real} \\
    \midrule
        Baseline	& ---	& 67.2	& 67.2 \\\hline
        Joint	    & 901k	& 67.0	& 59.9 \\
        LM	        & 19M	& {\bf 61.8}	& 56.3 \\
        Decoder	    & 20M	& 65.3	& 55.5 \\
        Encoder	    & 96M	& 65.8	& 58.3 \\
        All	        & 117M	& 64.8	& {\bf 50.4} \\
    \bottomrule
    \end{tabular}
\end{table}
First, we evaluate the WER performance of fine-tuning selected layers of the RNN-T model.
Table~\ref{tab:wer_selected_layers} shows the WER performance of the personalized models fine-tuned using either synthesized (TTS) or the actual (Real) speech data~\footnote{
Note that (Real) represents the best case scenario
where users correct all the recognition errors to achieve fully supervised training.}.
The table also shows the corresponding number of fine-tuned parameters.
Wiki-Names is a difficult dataset due to the presence of many name entities as well as the accented and disfluent speech. The baseline model has a rather high 67.2\% WER, making many errors for the names as well as short words (e.g. `a', `an' and `the').

Fine-tuning the entire model with the actual speech data achieves the lowest WER of 50.4\%. If we fine-tune a partial model, it is better to fine-tune the decoder compared to the encoder despite having about 5 times fewer parameters. This is expected because it is much easier for the LM component of the model to learn a new named entity. The additional gains from fine-tuning the entire model may come from learning the speakers' accents.
On the other hand, if we fine-tune the model with synthesized speech, we achieved the best WER performance of 61.8\% by updating the LM component only. This is again expected because fine-tuning the rest of the model would over-fit to the synthesized speech.

\subsection{Elastic Weight Consolidation}

\begin{table}[t]
    \centering
    \caption{Comparison of word error rate performance without ($\lambda=0$) and with ($\lambda=10^6$) EWC for different adaptation models (average over 4 speakers, one from each category).}
    \label{tab:wer_on_vs}
    \begin{tabular}{l|c|cc|cc}
    \toprule
        \multirow{3}*{\textbf{Model}} & 
        \multirow{3}*{\textbf{Epoch}} & 
        \multicolumn{4}{c}{\textbf{WER}} \\
        & & 
        \multicolumn{2}{c|}{\textbf{Voice Search}} &
        \multicolumn{2}{c}{\textbf{Wiki-Names}} \\
        & & 
        {\small $\lambda=0$} & 
        {\small $\lambda=10^6$} & 
        {\small $\lambda=0$} & 
        {\small $\lambda=10^6$} \\
    \midrule
        Baseline	& --- & 7.3 & 7.3 & 48.2 & 48.2 \\\hline
        \multirow{3}*{LM}
        & 5  & 7.3 & 7.3 & 47.6 & 46.9 \\
        & 10 & 7.6 & 7.5 & 40.0 & 41.1 \\
        & 15 & 8.3 & 7.8 & 35.9 & 36.5 \\\hline
        \multirow{3}*{All}
        & 5  & 8.6 & 8.3 & 42.3 & 43.1 \\
        & 10 & 10.1 & 8.8 & 35.2 & 32.7 \\
        & 15 & 11.4 & 9.7 & 32.0 & 28.7 \\
    \bottomrule
    \end{tabular}
\end{table}

Next, we study the effectiveness of Elastic Weight Consolidation (EWC), as described in Section~\ref{sec:ewc}, in handling a small amount of training data for speech personalization. We evaluate the performance of four personalized speech models (one speaker from each category) on a separate voice search test set, to understand how much the models have deviated from the baseline after personalization. Table~\ref{tab:wer_on_vs} shows the average performance of two types of personalized model (by fine-tuning either the LM component or the entire model) with different numbers of training epochs.
We compare training without ($\lambda=0$) and with ($\lambda=10^6$) EWC.
The baseline model achieves 7.3\% WER on the voice search test set.
As expected, WER increases as we personalize the models towards the Wiki-Names data with more training epochs. 
Without using EWC, we observe that the WER performance on Voice Search increases from 7.3\% to 8.3\% by fine-tuning the LM component and 11.4\% by fine-tuning the entire model. In this case, fine-tuning the LM component only offers a better trade-off between the two test sets.
For example, after 15 epochs, it achieves 35.9\% WER on Wiki-Names and 8.3\% on Voice Search. In comparison, fine-tuning the entire model for 10 epochs achieves a similar WER of 35.2\% on Wiki-Names, but with a higher WER of 11.4\% on Voice Search.
With elastic weight consolidation,
we were able to reduce the degradation on voice search. Surprisingly, we also achieve a better WER performance of 28.7\% on the Wiki-Names set when fine-tuning the entire model. The penalty term in EWC served as an effective regularizer to mitigate overfitting to a small amount of personalization data, and even assisted with learning the new data.

\subsection{Effect of TTS Engine}

\begin{table}[t]
    \centering
    \caption{Comparison of word error rate performance when using different TTS engines to generate labels.}
    \label{tab:wer_tts_accent}
    \begin{tabular}{l|c|c|c|c}
    \toprule
        \textbf{TTS} &
        \textbf{American} & 
        \textbf{Chinese} & 
        \textbf{Indian} & 
        \textbf{Italian} \\
    \midrule
        Baseline & 50.7 & 75.9 & 72.1 & 55.9 \\\hline
        en      & {\bf 45.2} & 70.4 & 68.1 & 49.6 \\
        zh      & 46.8 & {\bf 66.0} & 66.8 & 50.3 \\
        en-in   & 48.0 & 70.0 & {\bf 64.5} & 48.6 \\ 
        it	    & 46.6 & 68.6 & 66.5 & {\bf 46.5} \\
    \bottomrule
    \end{tabular}
\end{table}
Next, we study the effect of using different Text-To-Speech (TTS) engines for personalization.
We choose four TTS engines that match closely with the four name categories: English (en), Chinese (zh), Indian (en-in) and Italian (it). 
From the results in Table~\ref{tab:wer_tts_accent}, the choice of TTS engine does not matter so much compared to the gains over the baseline. Nevertheless, using matched TTS engines shows consistent improvements across all four name categories.

\subsection{Semi-supervised Learning}

\begin{table}[t]
    \centering
    \caption{Comparison of precision and recall (names) using the actual speech and different types of supervision. Personalization is achieved by fine-tuning the entire model (All). `Supervised' represents an ideal scenario.}
    \label{tab:semi_sup}
    \begin{tabular}{l|c|c|c|c|c}
    \toprule
        \multirow{2}*{\textbf{Supervision}} & 
        \multicolumn{2}{c|}{\textbf{Names}} & 
        \multicolumn{2}{c|}{\textbf{Non-names}} &
        \multirow{2}*{\textbf{WER}} \\
        & 
        \textbf{P} & 
        \textbf{R} &
        \textbf{P} & 
        \textbf{R} & \\ 
    \midrule
        Baseline                & 97.9 &  2.7 & 82.5 & 76.0 & 67.2 \\
        Unsupervised            & 96.7 &  3.5 & 83.8 & 62.2 & 69.1 \\
         + biasing              & 67.2 & 10.6 & 84.9 & 48.7 & 74.1 \\
        Semi-supervised & 68.8 & 68.2 & 91.1 & 61.3 & 63.8 \\
        Supervised           & 82.5 & 77.1 & 92.2 & 80.1 & 50.4 \\
    \bottomrule
    \end{tabular}
\end{table}
So far, we've assumed that correct transcriptions are available for training. 
However, training data for on-device personalization exists only on user devices, and attaining correct transcriptions on-device requires time-consuming effort from users. 
In a realistic scenario, users may not correct all the errors in the recognition outputs.
To better understand the impact of different amounts of user effort on personalization, we measure the breakdown of the precision and recall for the names of interest, as well as for the rest of the words in the sentence (non-names\footnote{{\em Non-names} may include other named entities that we do not track.}).
In Table~\ref{tab:semi_sup}, we compare using different types of supervision that require different amounts of user effort for training.
The baseline model performs poorly for the names, with a very low recall of 2.7\%. It rarely outputs the names, but it does so with a very high precision of 97.9\%.

For {\em unsupervised} training, we use the recognition output of the baseline model (67.2\% WER) as supervision for personalization. This requires no additional effort from the user to make corrections. However, there is no performance gain because the new names are not generated by the recognizer in the unsupervised transcripts.
Alternatively, if we use the biased recognition output as supervision, the name recall rate improved to 10.6\%. However, the precision for the names, the recall for the non-names as well as the overall WER performance degrade significantly because the model is learning from erroneous transcripts.

For {\em semi-supervised} training, we simulate the situation where users only correct the errors that matter (in this case the new named entities of interest). We do so by aligning the baseline recognition outputs with the reference and replace the words in the hypotheses that are aligned to the names in the references. For example, given the aligned reference and hypothesis in Table~\ref{tab:ref_hyp}, the name-corrected transcript will be `{\em Zhuge Dan was from Yangdu Zhuge}'. This assumes that the user corrects the missing `{\em Dan}' and the incorrectly recognized `{\em Yangdu}'. 
With a relatively small effort from the user (to correct only the names), we see a substantial improvement in recalling those names (3.5\% to 68.2\%, or 67.0\% relative improvement). However, the precision for the names and the recall for the non-names are still much worse than the baseline. The overall relative WER improvement is only about 5.1\% (from 67.2\% to 63.8\%).

\subsection{Biasing and TTS Data}

So far, we have considered fine-tuning the RNN-T model using full sentences.
Next, we also compare with two other personalization techniques that require only the names: 
1) biasing (as described in Section~\ref{sec:biasing});
and 2) fine-tuning the LM component using synthesized speech of the names only. 
\begin{table}[t]
    \centering
    \caption{Comparison of precision (P) and recall (R) for names/non-names and word error rate (WER). Supervised represents an ideal scenario.}
    \label{tab:with_biasing}
    \begin{tabular}{l|c|c|c|c|c}
    \toprule
        \multirow{2}*{\textbf{Method}} & 
        \multicolumn{2}{c|}{\textbf{Names}} & 
        \multicolumn{2}{c|}{\textbf{Non-names}} &
        \multirow{2}*{\textbf{WER}} \\ 
        & 
        \textbf{P} & 
        \textbf{R} &
        \textbf{P} & 
        \textbf{R} & \\ 
    \midrule
        Baseline	    & 100.0	& 2.4	& 85.7	& 59.6 & 70.0 \\
         + biasing      & 87.5	& 30.1	& 87.5	& 63.7 & 63.5 \\\hline
        TTS (names)	    & 91.6  & 13.4  & 87.2  & 60.3 & 69.1 \\
         + biasing	    & 82.3  & 34.5  & 90.1  & 59.7 & 66.7 \\\hline
        TTS (sentences)	& 90.1  & 22.5  & 88.2  & 64.2 & 65.2 \\
         + biasing	    & 76.9  & 48.6  & 91.4  & 64.5 & 62.2 \\\hline
        Semi-supervised  & 75.5  & 52.5  & 92.8  & 51.4 & 68.3 \\
         + biasing	    & 63.0  & 64.4  & 93.9  & 51.8 & 66.9 \\\hline
        Supervised	        & 88.1  & 65.0  & 93.2  & 71.7 & 54.8 \\
         + biasing	    & 80.1  & 73.5  & 94.0  & 71.5 & 53.8 \\
    \bottomrule
    \end{tabular}
\end{table}
Table~\ref{tab:with_biasing} shows the comparison of word error rate as well as the precision and recall performances of four personalization techniques. For this experiment, we use a slightly different decoding setup that includes end-pointing and supports biasing. The results in Table~\ref{tab:with_biasing} are not directly comparable with the earlier results in the paper, but are self consistent. 

Our results show that training with synthesized sentences achieves better results than training with synthesized names only. This is not surprising because the model sees the names in different linguistic contexts during training. Moreover, training with the actual speech data performs better than using synthesized data because the model learns the user-specific pronunciation of the names. 

We also observe that biasing can be applied to fine-tuned models and obtain consistent further improvements. With biasing, the recall rate for the names improve as the user provides more supervision signals:
30.1\% (biasing only) $\rightarrow$
48.6\% (with TTS sentences) $\rightarrow$
64.4\% (actual speech with name-corrected transcripts) $\rightarrow$
73.5\% (fully supervised).
However, biasing increases the number of false alarms for the names, which results in lower precision for names.

\subsection{On-device Benchmark}

\begin{table}[t]
    \centering
    \caption{Benchmark Results of training memory and speed on a Pixel 3 mobile phone with different batch sizes.}
    \label{tab:benchmark}
    \begin{tabular}{l|c|c}
    \toprule
        \textbf{Batch Size} & 
        \textbf{Memory (GB)} & 
        \textbf{Epoch Time (minutes)} \\
    \midrule
        1 & 1.5 & 50 \\
        5 & 1.6 & 22 \\
        10 & 1.5 & 14 \\
        20 & 1.8 & 10 \\
    \bottomrule
    \end{tabular}
\end{table}
Finally, we ran benchmark experiments to measure the memory and speed performances for on-device training. Training was done on a Pixel 3 mobile phone using 50 utterances from one speaker over 1 epoch.
The benchmark results are shown in Table~\ref{tab:benchmark}, comparing different batch sizes.
The batch size does not affect the memory usage much.
Training consumes between 1.5 -- 1.8 GB memory with batch sizes 1, 5, 10 and 20.
On the other hand, increasing the batch size significantly improves the throughput. With batch size of 1, the time needed to perform one epoch of training is 50 minutes. By increasing the batch size to 20, the training time per epoch reduces by 5 times to 10 minutes.
Therefore, training on device for 15 epochs with 5 utterances per mini batch (which is the configuration used for the experiments in this paper) will take about 5.5 hours. Increasing the batch size to 20 will reduce the training time to 2.5 hours.
This fits reasonably well within one typical training session when the phone is idle.
Moreover, personalization can also be performed over multiple training sessions to progressively improve the model as more user data become available. 

\section{Conclusions}
\label{sec:conclusions}

This paper investigates the effectiveness of several techniques for personalizing end-to-end speech recognition models on device by learning new named entities.
We find that it is possible to fine-tune the entire model with 117 million parameters using a small learning rate and early stopping. Elastic weight consolidation can be used to mitigate overfitting, and achieves a 10.3\% relative word error rate improvement on the Wiki-Names dataset.
Since supervised on-device training would rely on the user to provide labeled data, 
we compare training with differing amounts of supervision (and corresponding user effort).
We propose using {\em keyword-dependent} precision and recall to measure vocabulary acquisition performance.
Our results show that biasing combined with synthesized data of sentences can improve the recall of new named entities from 2.4\% to 48.6\% without the users providing speech inputs and/or correcting the recognition outputs. The name recall rate can be further improved to 64.4\% and 73.5\%
if the user corrects only the errors for the names ({\em semi-supervised}) and all the errors ({\em supervised}), respectively.

\bibliographystyle{IEEEbib}
\bibliography{refs}

\end{document}